\documentclass[10pt,fleqn]{article}
\usepackage{amssymb}

\newcommand{\name}[1]{\begin{flushleft}
                       \LARGE \bf #1
                       \end{flushleft}\vspace{-3mm}}

\newcommand{\Author}[1]{\begin{flushleft}
                       \it #1 \end{flushleft}}

\newcommand{\Adress}[1]{\begin{flushleft}
                       \it #1 \end{flushleft}}

\newcommand{\be}{\begin{equation}}
\newcommand{\ee}{\end{equation}}
\newcommand{\ba}{\hspace*{-5pt}\begin{array}}
\newcommand{\ea}{\end{array}}
\newcommand{\p}{\partial}
\newcommand{\ds}{\displaystyle}

\newcommand{\pbf}[1]{\mbox{\mathversion{bold}$#1$}}

\begin{document}

\name{On a motion equation for two particles\\
 in relativistic quantum mechanics}

\medskip

\noindent{published in {\it Lettere al Nuovo Cimento},  1974,
{\bf 10}, N~4, P. 163--167.}

\Author{Wilhelm I. FUSHCHYCH\par}

\Adress{Institute of Mathematics of the National Academy of
Sciences of Ukraine, \\ 3 Tereshchenkivska  Street, 01601 Kyiv-4,
UKRAINE}

\noindent {\tt URL:
http://www.imath.kiev.ua/\~{}appmath/wif.html\\ E-mail:
symmetry@imath.kiev.ua}

\bigskip

\noindent
Breit [1] was the first who proposed to describe the motion for two relativistic particles
by means of a semi-relativistic Dirac-type equation. The wave function of this equation
has sixteen components. The possibility of covariant description of a system of par\-tic\-les
interacting m quantum mechanics was proved by Thomas and Bakamjian~[2]
and Foldy~[3]. In quantum field theory the two-body problem is described by means of the
Bethe--Salpeter equation or the Logunov--Tavkhelidze--Kadyshevsky equations~[4].

The purpose of the present note is to propose, in the framework of relativistic quantum
mechanics, a new Poincar\'e-invariant equation for two particles with masses $m_1$, $m_2$
and spin $s_1=s_2=\frac 12$.
It is a first-order linear differential equation for the eight-component wave function.
With the help of this equation the description of the motion of two-particle systems is
reduced to the description of one-particle systems in the $(1+6)$-dimensional
Minkowski space which can be in two spin states ($s=0$ or $s=1$).

At first we derive the equation for two noninteracting particles. To this end we shall pass from
the momenta of two particles $\pbf{p}_1$, $\pbf{p}_2$
to the new canonical variables
\[
\pbf{P}=(P_1,P_2, P_3) =\pbf{p_1}+\pbf{p}_2, \qquad \pbf{K}=(K_1,K_2,K_3).
\]
The connection between the variables  $\pbf{K}$ and $\pbf{p}_1$, $\pbf{p}_2$
is rather complicated (see, e.g., [5, 6])
and we do not equate it here. The total energy of the two-particle system in the variables
$\pbf{P}$ and $\pbf{K}$ has for our discussion a very convenient structure~[5, 6]
\be
E=(\pbf{P}^2+M^2)^{1/2}, \qquad M=\left(m_1^2+\pbf{K}^2\right)^{1/2}
+\left(m_2^2+\pbf{K}^2\right)^{1/2}.
\ee
The square energy for the case when $m_1=m_2\equiv \frac 12 m$ takes the very simple form
\be
E^2=p_a^2+p_{a+3}^2 +m^2, \qquad p_a\equiv P_a, \qquad p_{a+3}\equiv 2K_a, \qquad
a=1,2,3.
\ee
The square root from this expression is the equation for two particles
\be
i\frac{\p \Psi(t,x_1,x_2,\ldots,x_6)}{\p t} ={\mathcal H}(\hat p_1, \hat p_2, \ldots, \hat p_6)
\Psi(t,x_1,x_2,\ldots,x_6),
\ee
where
\be
\ba{l}
\ds {\mathcal H}(\hat p_1, \hat p_2, \ldots, \hat p_6)=\Gamma_0 \Gamma_a \hat p_a +
\Gamma_0 \Gamma_{a+3} \hat p_{a+3} +\Gamma_0 m,
\vspace{2mm}\\
\ds  \hat p_a =-i\frac{\p}{\p x_a}, \qquad \hat p_{a+3}=-i\frac{\p}{\p x_{a+3}},
\ea
\ee
the $8\times 8$ matrices $\Gamma_0$, $\Gamma_a$, $\Gamma_{a+3}$
obey a Clifford algebra, and has such a representation:
\be
\Gamma_0 =\sigma_3 \otimes 1, \qquad \Gamma_a =2i\sigma_2 \otimes s_a,
\qquad \Gamma_{a+3}=2i\sigma_1 \otimes \tau_a,
\ee
$$
\ba{l}
\ds s_1 =\frac 12\left( \begin{array}{cccc}
0 & 1 & 0 & 0 \\ 1 & 0 & 0 &0 \\ 0 & 0 & 0 & -i \\ 0 & 0 & i &0 \end{array}\right), \qquad
s_2=\frac 12\left( \begin{array}{cccc}
0 & 0& 1& 0 \\ 0& 0 & 0 &i\\ 1& 0 & 0 & 0 \\ 0 & -i& 0 &0 \end{array}\right),
\vspace{2mm}\\
\ds s_3 =\frac 12\left( \begin{array}{cccc}
0 & 0 & 0 & 1 \\ 0 & 0 & -i &0 \\ 0 & i & 0 & 0 \\ 1 & 0 & 0 &0 \end{array}\right),
\qquad
\ds \tau_1 =\frac 12\left( \begin{array}{cccc}
0 & -1 & 0 & 0 \\ -1 & 0 & 0 &0 \\ 0 & 0 & 0 & -i \\ 0 & 0 & i &0 \end{array}\right),
\vspace{2mm}\\
\ds \tau_2=\frac 12\left( \begin{array}{cccc}
0 & 0& -1& 0 \\ 0& 0 & 0 &i\\ -1& 0 & 0 & 0 \\ 0 & -i& 0 &0 \end{array}\right), \qquad
\tau_3 =\frac 12\left( \begin{array}{cccc}
0 & 0 & 0 & -1 \\ 0 & 0 & -i &0 \\ 0 & i & 0 & 0 \\ -1 & 0 & 0 &0 \end{array}\right),
\ea
$$
The $\sigma_a$ are the Pauli matrices.

The two-particle equation (3) will be defined completely in that case if we de\-ter\-mi\-ne both
the Hamiltonian and the Poincar\'e generators~[7].
The generators of the $P_{1,3}$ group on $\{\Psi\}$ have such a form:
\be
\ba{l}
\ds P_0={\mathcal H}(\hat p_1,\ldots, \hat p_6)=\Gamma_0 \Gamma_A\hat p_A+\Gamma_0 m,
\qquad P_a=p_a, \qquad A=1,2,\ldots,6,
\vspace{2mm}\\
\ds J_{ab}=M_{ab}+m_{ab}+S_{ab}, \qquad a,b=1,2,3,
\vspace{2mm}\\
\ds J_{0a}=t p_a -\frac 12 (x_a {\mathcal H}+{\mathcal H} x_a)-\frac{{\mathcal H}}{
\sqrt{{\mathcal H}^2}} \frac{(S_{ab}^{(2)}+m_{ab})p_b}
{\sqrt{{\mathcal H}^2} +M},
\ea
\ee
where
\be
\ba{l}
\ds M_{ab}\equiv\hat x_a \hat p_b -\hat x_b \hat p_a, \qquad
m_{ab}\equiv \hat x_{a+3} \hat p_{b+3}-\hat x_{b+3} \hat p_{a+3},
\qquad S_{ab}=S_{ab}^{(1)}+S_{ab}^{(2)},
\vspace{2mm}\\
\ds S_{ab}^{(1)} =\frac i4 (\Gamma_a \Gamma_b -\Gamma_b \Gamma_a),
\qquad S_{ab}^{(2)} =\frac i4 (\Gamma_{a+3} \Gamma_{b+3} -\Gamma_{b+3}\Gamma_{a+3}),
\vspace{2mm}\\
\ds \left[\hat x_a, \hat p_b\right]_-=i\delta_{ab}, \qquad
\left[\hat x_{a+3}, \hat p_{b+3}\right]_-=i\delta_{ab},
\vspace{2mm}\\
\ds \left[\hat x_a,\hat x_b\right]_-=\left[\hat x_a, \hat x_{a+3}\right]_-=
\left[\hat x_{a+3}, \hat x_{b+3}\right]_-=0,
\qquad \left[ \hat x_a, \hat p_{b+3}\right]_-=\left[\hat x_{a+3}, \hat p_b\right]_-=0.
\ea\hspace{-12.03pt}
\ee
It can be immediately verified that the operators (6) satisfy the Poincar\'e algebra.
It follows that eq. (3) is Poincar\'e invariant. If we perform the unitary transformation
\be
U=\frac{(E+M+\Gamma_c p_c)(M+m+\Gamma_{c+3}p_{c+3})}{2\{ME(E+m)(M+m)\}^{1/2}}
\ee
on the operators (6), then we obtain
\be
\ba{l}
\ds P_0^c =UP_0U^\dag =\Gamma_0 E, \qquad P_a^c=p_a, \qquad
J_{ab}^c=UJ_{ab}U^\dag =J_{ab},
\vspace{2mm}\\
\ds J_{0a}^c =t p_a -\frac 12 (x_a P_0^c+P_o^c x_a)-\Gamma_0 \frac{m_{ab} p_b +S_{ab}p_b}
{E+M}.
\ea
\ee
The transformed generators (9) have canonical form [2, 3].
The position operators $X_a$ and $X_{a+3}$ on a set $\{\Psi\}$ look like
\be
X_a =U^\dag x_a U=x_a +\frac{S_{ab}^{(1)} p_b}{E(E+M)}+i\left( \frac{\Gamma_a}{2E}-
\frac{p_a \Gamma_cp_c}{2E^2 (E+M)} \right)
\frac{m+\Gamma_{c+3}p_{c+3}}{M},
\ee
\be
\ba{l}
\ds X_{a+3}=U^\dag x_{a+3} U=x_{a+3}+\frac{S_{a+3\, b+3}^{(2)} p_{b+3}}{M(M+m)}
+\frac{i\Gamma_{a+3}}{2M} -
\vspace{2mm}\\
\ds \qquad -i\frac{p_{a+3} \Gamma_{c+3} p_{c+3}}{2M^2(M+m)}-
i\frac{p_{a+3}}{2E^2 M^2} \Gamma_c p_c (m+\Gamma_{c+3} p_{c+3}).
\ea
\ee

An interaction Hamiltonian for two particles, in the absence of external fields, can have the form
\be
{\mathcal H}=\Gamma_0 \Gamma_A p_A+\Gamma_0\{m^2 +V(r)\}^{1/2},
\ee
where  $V(r)$ is an arbitrary function depending on $r\equiv \sqrt{x_{c+3}^2}$.
In the special case when $V(r)=e^4/r^2$ the interaction Hamiltonian can be written as
\be
{\mathcal H}=\Gamma_0^{(16)}\Gamma_A^{(16)} p_A+\frac{e^2}{r} \Gamma_0^{(16)}
\Gamma_7^{(16)} +\Gamma_0 m,
\ee
where the $16\times 16$ matrices $\Gamma_0^{(16)}$, $\Gamma_A^{(16)}$,
$\Gamma_7^{(16)}$ satisfy a Cifford algebra. An external elec\-t\-ro\-mag\-ne\-tic field is
introduced in eq. (3) in the following way:
\[
p_a \to \pi_a =p_a -e{\mathcal A}_a (t,x_1,x_2,x_3),
\quad \! p_{a+3}\to \pi_{a+3}=p_{a+3}-e{\mathcal A}_{a+3}(t,x_4,x_5,x_6).
\]

An extraction of the positive solutions from eq. (3) is realized by means of the subsidiary
condition
\[
\left( 1-\frac{{\mathcal H}}{\sqrt{{\mathcal H}^2}} \right)\Psi =0 \qquad \mbox{or}
\qquad \left( 1 -\frac{\Gamma_\mu p^\mu}{\sqrt{p_\mu^2}}\right)\Psi=0, \qquad \mu=0,1,2,\ldots,6.
\]
It is evident that these conditions are invariant under the Poincar\'e group.

It should be noted that the function $V(r)$ may be of arbitrary form, therefore the relative velocity
${\mathcal V}_{a+3}$,
\be
\hat {\mathcal V}_{a+3}\Psi\equiv -i[X_{a+3}, {\mathcal H}] \Psi ={\mathcal V}_{a+3}\Psi,
\ee
with respect to the centre-of-mass may be arbitrary. To do ${\mathcal V}_{a+3}$
smaller than the photon velocity it is necessary to impose the condition
\[
{\mathcal V}_{a+3}^2 ={\mathcal V}_4^2 +{\mathcal V}_5^2 +{\mathcal V}_6^2<1.
\]
These questions will be considered in more detail in another paper.

Finally we shall find the equation for two particles with mass $m_1\not= m_2$.
Let us, with Kadyshevsky et al.~[8],  represent $M$ in such a form
\be
M=\frac{m_1+m_2}{\sqrt{m_1m_2}} (m_1m_2 +{\pbf{K}'}^2)^{1/2},
\ee
where
\be
{\pbf{K}'}^2 =-m_1m_2 +\frac{m_1m_2}{(m_1+m_2)^2} \left( \sqrt{m_1^2 +\pbf{K}^2}+
\sqrt{m_2^2 +\pbf{K}^2}\right)^2.
\ee
In the variables $\pbf{P}$ and $\pbf{K}'$ formula (2) can be rewritten as
\be
E^2 =\pbf{P}^2 +\frac{(m_1+m_2)^2}{m_1m_2} {\pbf{K}'}^2 +(m_1+m_2)^2.
\ee
It follows that the equation of motion for the two particles is
\be
\ba{l}
\ds i \frac{\p \Psi(t,x_1,\ldots,x_6)}{\p t}=\Biggl\{ \Gamma_0 \Gamma_a \hat p_a+
\frac{m_1 +m_2}{\sqrt{m_1m_2}} \Gamma_0 \Gamma_{a+3} \hat p_{a+3} +
\vspace{2mm}\\
\ds \qquad \qquad \qquad +(m_1+m_2) \Gamma_0\Biggr\} \Psi(t,x_1,\ldots,x_6),
\vspace{2mm}\\
\ds \hat p_a =-i\frac{\p}{\p x_a}, \qquad \hat p_{a+3} \equiv \hat{\pbf{K}}'_a=-i
\frac{\p}{\p x_{a+3}}.
\ea
\ee
In this equation $\Psi$ is also an eight-component function.

\smallskip

I wish to thank A.G. Nikitin, A.L.~Grishchenko for useful comments and N.V.~Hna\-tjuk
for helpful rending the manuscript.

\medskip

\begin{enumerate}
\footnotesize

\item Breit G., {\it Phys. Rev.}, 1929, {\bf 34}, 553.

\item Bakamjian B.,  Thomas L.H., {\it Phys. Rev.}, 1953, {\bf 92}, 1300.

\item Foldy L.L., {\it Phys. Rev.}, 1961, {\bf 122}, 289.

\item Logunov A.A., Tavkhelidze A.N., {\it  Nuovo Cimento}, 1963, {\bf 29}, 380;\\
Kadyshevsky V.G.., {\it Nucl. Phys.}, 1968, {\bf 36}, 125.

\item Fong R., Sucher J., {\it J. Math. Phys.}, 1964, {\bf 5}, 456.

\item Osborn H., {\it Phys. Rev.}, 1968, 176, 1514.

\item Fushchych W.I., {\it Lett. Nuovo Cimento}, 1973, {\bf 6}, 133, {\tt quant-ph/0206105}.

\item Kadyshevsky V.G., Mateev M.D., Mir-Kasimov R.M., {\it J. Nucl. Phys.}, 1970,
{\bf 11}, 692 (in Russian).

\end{enumerate}
\end{document}